\documentclass[12pt,onecolumn,journal]{IEEEtran}

\usepackage[final]{graphicx}
\usepackage[reqno]{amsmath}
\usepackage{amssymb}

\newfont{\bbb}{msbm10 scaled 500}

\newfont{\bb}{msbm10 scaled 1100}

\newcommand{\NN}{\mbox{\bb N}}

\newcommand{\vv}{{\bf v}}

\newcommand{\Hm}{{\bf H}}

\newcommand{\Vm}{{\bf V}}
\newcommand{\Xm}{{\bf X}}
\newcommand{\Ym}{{\bf Y}}
\newcommand{\Zm}{{\bf Z}}

\newcommand{\Cc}{{\cal C}}

\newcommand{\Kc}{{\cal K}}
\newcommand{\Lc}{{\cal L}}
\newcommand{\Mc}{{\cal M}}
\newcommand{\Nc}{{\cal N}}

\newcommand{\Sc}{{\cal S}}

\newcommand{\Wc}{{\cal W}}

\newtheorem{theorem}{Theorem}
\newtheorem{corollary}[theorem]{Corollary}
\newtheorem{lemma}[theorem]{Lemma}

\IEEEoverridecommandlockouts

\author{
 \authorblockN{Onur Ozan Koyluoglu, Hesham El Gamal,
 Lifeng Lai, and H. Vincent Poor
 \thanks{Onur Ozan Koyluoglu and Hesham El Gamal
 are with the Department of Electrical and Computer Engineering,
 The Ohio State University, Columbus, OH 43210 USA.
 Hesham El Gamal also serves as the Director for the
 Wireless Intelligent Networks Center (WINC), Nile University, Cairo, Egypt.
 Email: \{koyluogo,helgamal\}@ece.osu.edu.}
 \thanks{Lifeng Lai and H. Vincent Poor
 are with the Department of Electrical Engineering,
 Princeton University, Princeton, NJ 08544, USA.
 Email: \{llai,poor\}@princeton.edu.}
 \thanks{This work is submitted to IEEE Transactions on Information Theory
 on October 7, 2008.}
 \thanks{The material in this paper was presented in part at the IEEE 
 International Symposium on Information Theory, Toronto, ON in July 2008.
 This research was supported in
 part by the US NSF under Grants CCF-07-28762, CCF-03-46887,
 ANI-03-38807, CNS-06-25637, and CCF-07-28208.}}}

\title{Interference Alignment for Secrecy}

\begin{document}
\maketitle


\begin{abstract}
This paper studies the frequency/time selective $K$-user Gaussian
interference channel with secrecy constraints. Two distinct models,
namely the interference channel with confidential messages and the
one with an external eavesdropper, are analyzed. The key difference
between the two models is the lack of channel state information
(CSI) about the external eavesdropper. Using interference alignment
along with secrecy pre-coding, it is shown that each user can
achieve non-zero secure Degrees of Freedom (DoF) for both cases.
More precisely, the proposed coding scheme achieves
$\frac{K-2}{2K-2}$ secure DoF {\em with probability one} per user in
the confidential messages model. For the external eavesdropper
scenario, on the other hand, it is shown that each user can achieve
$\frac{K-2}{2K}$ secure DoF {\em in the ergodic setting}.
Remarkably, these results establish the {\em positive impact} of
interference on the secrecy capacity region of wireless networks.
\end{abstract}


%


\section{Introduction}\label{sec:Introduction}

The wiretap channel was introduced by Wyner~\cite{Wyner:Wiretap75},
in which the eavesdropper is assumed to have access to a degraded
version of the intended receiver's signal. This pioneering work was
later generalized to cover the non-degraded
scenario~\cite{Csiszar:Broadcast78} and the Gaussian
channel~\cite{Cheong:The78}. However, these results show that the
secrecy capacity saturates in the high signal-to-noise ratio (SNR)
regime implying a {\bf vanishing} value for the secure degrees of
freedom (DoF).

Recently, there has been a growing interest in the analysis and
design of secure wireless communication networks based on
information theoretic principles. For example, the secrecy capacity
of relay networks was studied in~\cite{Oohama:Coding01,Lai:The},
while the fundamental limits of the wiretap channel with feedback
were analyzed by~\cite{Lai:The07}. On the other hand, the multiple
access and broadcast channels with secrecy constraints were
investigated in~\cite{Tekin:The08,Liang:Generalized,Khisti:Secure}.
Finally, the role of multiple antennas in enhancing the secrecy
capacity was established in~\cite{Khisti:The07,Oggier:The} and the
positive impact of fading on secrecy capacity was revealed in
\cite{Liang:Secure, Gopala:On}.

Here, the frequency/time selective $K$-user Gaussian interference
channel with secrecy constraints is considered. Without the secrecy
constraints, it has been recently shown that a $\frac{1}{2}$ degrees
of freedom (DoF) per orthogonal dimension is achievable for each
source-destination pair in this network~\cite{Cadembe:Interference}.
The achievability of this result was based on the \emph{interference
alignment} technique (see also~\cite{Maddah-Ali:Communication08}),
by which the interfering signals are aligned to occupy a subspace
orthogonal to the one spanned by the intended signal at each receiver.
However, the impact of secrecy constraints on the degrees of freedom in
this model has not been fully characterized. In fact, to the best of
our knowledge, the only relevant prior works are the study of the
two-user discrete memoryless interference channels with confidential
messages~\cite{Liu:Discrete06,Liu:Discrete,Liang:Cognitive07} and
the one with an external eavesdropper~\cite{Koyluoglu:Onthe08}. The
frequency selective interference channel adopted in the present
paper is, however, fundamentally different from these {\em
memoryless} models.

We consider two distinct network models, namely 1) the interference
channel with confidential messages and 2) the one with an external
eavesdropper. In the first scenario, one needs to ensure the {\em
confidentiality} of each message from all non-intended receivers in
the network. Since all users are assumed to belong to the same
network, one can assume the availability of channel state
information (CSI) while designing the secrecy coding scheme. Towards
this goal, we employ an interference alignment scheme along with
secrecy pre-coding at each transmitter. Intuitively, the
interference alignment scheme has two effects on each receiver $i$:
1) it aligns the signals from transmitters $k\neq i$ to a small
dimensionality subspace, and 2) it assigns the signal from
transmitter $i$ to the orthogonal subspace. Hence, while the signal
from its own transmitter is {\em received cleanly}, the signals from
other transmitters {\em are mixed together}. Our secrecy pre-coding
takes advantage of the phenomenon to ensure that the resulting
multiple access channel (from the $K-1$ interfering users) does not
reveal any useful information about each non-intended message. This
way, we show that $\frac{1}{4}$ {\bf secure} DoF per orthogonal
dimension is achievable for each user in the three-user Gaussian
interference channel with confidential messages. We then generalize
our results to the $K$-user Gaussian interference channel showing
that each user can achieve $\frac{K-2}{2K-2}$ secure DoF. In the
second scenario, we study the external eavesdropper model where the
fundamental challenge is the lack of channel state information (CSI)
about the links connected to it. Despite this fact, it is shown that
$1/2-1/K$ DoF per user is achievable in the ergodic setting. This
result provides further evidence on the diminishing gain resulting
from knowing the instantaneous CSI of the eavesdropper {\em
a-priori}. Interestingly, by comparing our results with those
obtained for the point-to-point case~\cite{Liang:Secure, Gopala:On},
one can see the {\bf positive impact} of interference on the secrecy
capacity region of wireless network. The underlying idea is that the
coordination between several source-destination pairs allows for
{\em hiding} the secret messages in the background interference.

The remainder of the paper is organized as follows. In Section II,
the system model and notation are introduced. Section III is devoted
to the interference channel with confidential messages. The analysis
for the external eavesdropper scenario is detailed in Section IV.
Finally, we offer some concluding remarks in Section V. The
technical results needed to develop our proofs are collected in the
appendices to enhance the flow of the paper.


\section{System Model}
\label{sec:Model}

\subsection{The Confidential Messages Scenario}

We consider a frequency selective wireless network comprised of
$K$ transmitter-receiver pairs where the  $i^{th}$ receiver output
at time $t\in\{1,\cdots,n\}$ and frequency slot
$f\in\{1,\cdots,F\}$ is given by
\footnote{In this paper, matrices are represented
with bold capital letters ($\Xm$) and vectors are denoted as bold
capital letters with bars or tildes (for example, $\bar{\Xm}$ and
$\tilde{\Xm}$). We define $\Kc\triangleq \{1,\cdots,K\}$ and
denote $\Xm_\Sc \triangleq \{\Xm_k|k\in\Sc\}$ for
$\Sc\subset\Kc$. A zero-mean circularly symmetric complex Gaussian
random variable with variance $\sigma^2$ is denoted by
$\Cc\Nc(0,\sigma^2)$.}

\begin{equation}\label{eq:model}
Y_i (f,t) = \sum\limits_{k=1}^K h_{ik} (f) X_k (f,t) + Z_i (f,t).
\end{equation}
Here, $X_k (f,t)$ is the transmitted symbol of user $k$ at
frequency slot $f$ during time $t$, and $Z_i (f,t)\sim \Cc\Nc(0,1)$
is the additive white Gaussian noise at receiver $i$. We assume that
the channel coefficients are randomly generated according to a continuous
distribution and are fixed during the communication period. We
also assume that the channel coefficients are known at every node
in the network. The network model is provided in Fig.~$1$.

Using the extended channel notation
of~\cite{Cadembe:Interference}, the $i^{th}$ received vector
during time slot $t$ can be written as
\begin{equation}
\bar{\Ym}_i (t) = \sum\limits_{k=1}^K\Hm_{i,k} \bar{\Xm}_k (t) +
\bar{\Zm}_i (t).
\end{equation}
Here, $\Hm_{i,k}$ is the $F \times F$ diagonal matrix of channel
coefficients from transmitter $k$ to receiver $i$ whereas
$\bar{\Ym}_i (t)=[Y_i(1,t),\cdots, Y_i(F,t)]^T$,
$\bar{\Zm}_i(t)=[Z_i(1,t),\cdots, Z_i(F,t)]^T$, and $\bar{\Xm}_k
(t)=[X_k(1,t),\cdots, X_k(F,t)]^T$ are $F \times 1$ column vectors.

We assume that each source $k\in\Kc$ has a message $W_k$ which must
be transmitted in secrecy from the remaining $K-1$ receivers.
Therefore, our $(n,F,M_1,\cdots,M_K)$ secret codebook has the
following components:

$1$) The secret message set $\Wc_k=\{1,\cdots,M_k\}$.

$2$) Encoding functions $f_k(.)$ which map the secret messages to
the transmitted symbols, i.e., $f_k:w_k\to (\bar{\Xm}_k
(1),\cdots,\bar{\Xm}_k (n))$ for each $w_k\in\Wc_k$. At encoder $k$,
each codeword is designed according to the transmitter's average
long-term power constraint $\rho$, i.e.,
$$\frac{1}{nF}\sum\limits_{f=1}^F \sum\limits_{t=1}^n
(X_k(t,f))^2\leq \rho.$$

$3$) Decoding functions $\phi_k(.)$ at receivers $k\in\Kc$ which map
the received symbols to estimates of the messages:
$\phi_k(\Ym_k)=\hat{W}_k$ where $\Ym_k=\{\bar{\Ym}_k
(1),\cdots,\bar{\Ym}_k (n)\}$.

The reliability of the transmission of user $k$ is measured
by the probability of error
\begin{eqnarray}
P_{e,k}=\frac{1}{\prod\limits_{i=1}^{K}M_i}\:\:\sum\limits_{(w_1,\cdots,w_K)
\:\in\:\Wc_1\times\cdots\times\Wc_K}\text{Pr}
\left\{\phi_k(\Ym_k)\neq w_k | (w_1,\cdots,w_K)
\textrm{ is sent}\right\},\nonumber
\end{eqnarray}
whereas the secrecy level is measured by the normalized equivocation
defined as follows~\cite{Wyner:Wiretap75,Cheong:The78}: For receiver
$i$, the equivocation for each subset of messages $W_\Sc$,
$\Sc\subset\Kc-i$, is
$$\Delta_{\Sc,i}\triangleq\frac{H\left(W_\Sc|\Ym_i\right)}{H\left(W_\Sc\right)}.$$

We say that the rate-equivocation tuple $(R_1,\cdots,R_K,d)$ is achievable
for the Gaussian interference channel with
confidential messages, if, for any given $\epsilon > 0$, there exists an
$(n,F,M_1,\cdots,M_K)$ secret codebook such that,
\begin{eqnarray}\label{eq:secrecy}
\frac{1}{nF} \log_2 M_k &=& R_k,
\:\forall k\in\Kc \nonumber,\\
\max\{P_{e,1},\cdots,P_{e,K}\}&\leq& \epsilon,
\end{eqnarray} and
\begin{eqnarray}
\Delta_{\Sc,i} &\geq& d-\epsilon, \: \forall i\in\Kc,\: \forall
\Sc\subset\Kc-i \nonumber,
\end{eqnarray}
where a symmetric secrecy notion for each user in the network is
used. We also say that the symmetric degrees of freedom (per
orthogonal frequency-time slot) of $\eta$ is achievable with perfect
secrecy, if the rate-equivocation tuple $(R_1=R,\cdots,R_K=R,d=1)$
is achievable and
$$\eta=\lim
\limits_{\rho\to\infty} \frac{R}{ \log(\rho)}.$$

\subsection{The External Eavesdropper Scenario}

In this model, we assume the existence of an external eavesdropper
who observes the signals of the $K$ sources (see Fig.~$2$). We
consider an ergodic setting where the channel gains are fixed during
a block of $n_1$ symbol times and then randomly change to another
value for the next block. Hence, transmission time of $n$ time slots
is divided into $B$ fading blocks with $n=n_1B$. We denote the
received signals at receiver $i\in\{1,\cdots,K,e\}$ using the
extended channel notation as follows
\begin{equation}
\bar{\Ym}_i(j+(b-1)n_1)=\sum\limits_{k=1}^K \Hm_{i,k}(b)
\bar{\Xm}_{k}(j+(b-1)n_1) + \bar{\Zm}_{i}(j+(b-1)n_1),
\end{equation}
where $b\in\{1,\cdots,B\}$ denotes the fading block $b$, $j\in
\{1,\cdots,n_1\}$ denotes the $j^\textrm{th}$ time instant of the
corresponding fading block, $\Hm_{i,k}(b)$ is the $F\times F$
diagonal matrix of channel coefficients between transmitter $k$ and
receiver $i$ during fading block $b$, and
$\bar{\Xm}_{k}(j+(b-1)n_1)$ is the transmitted vector of user $k$
at $j^\textrm{th}$ symbol of the $b^\textrm{th}$ fading block. We
further define $\Hm\triangleq\{\Hm_{i,k}(b): i,k\in\Kc,
b\in\{1,\cdots,B\} \}$ and $\Hm_e\triangleq\{\Hm_{e,k}(b): k\in\Kc,
b\in\{1,\cdots,B\} \}$. We assume that $\Hm$ is known at all the
nodes in the network, whereas $\Hm_e$ is known only at the
eavesdropper (only the statistical knowledge about the eavesdropper
CSI is available to the network users). The channel coefficients are
i.i.d. samples of a zero-mean unit variance complex Gaussian
distribution.

The components of the secrecy codebook remain as before with the
exception that each transmitter must secure its own message
\emph{only} from the external eavesdropper. Accordingly, we modify
the secrecy requirement by considering the normalized equivocation
seen by the eavesdropper. We denote the observation at the
eavesdropper as $\Ym_e=\{\bar{\Ym}_e (1),\cdots,\bar{\Ym}_e (n)\}$,
in which $\bar{\Ym}_e (t)$ is defined similarly as $\bar{\Ym}_i (t)$
for $t=1,\cdots, n$. Therefore, the normalized equivocation for a
subset of messages $\Sc\subset\Kc$ is given by
$$\Delta_{\Sc}\triangleq
\frac{H\left(W_\Sc|\Ym_e,\Hm,\Hm_e\right)}{H\left(W_\Sc\right)}.$$

We say that the rate-equivocation
tuple $(R_1,\cdots,R_K,d)$ is achievable
for the Gaussian interference channel with
an external eavesdropper, if, for any given $\epsilon>0$, there
exits an $(n,F,M_1,\cdots,M_K)$ secret codebook such that
\begin{eqnarray}\label{eq:secrecy2}
\frac{1}{nF} \log_2 M_k &=& R_k,
\: \forall k\in\Kc, \nonumber\\
\max\{P_{e,1},\cdots,P_{e,K}\}&\leq& \epsilon,
\end{eqnarray}
and
\begin{eqnarray}
\Delta_{\Sc} &\geq& d - \epsilon
,\: \forall \Sc\subset\Kc \nonumber.
\end{eqnarray}

It then follows that the symmetric DoF with perfect secrecy is
defined along the same lines as in the previous section.


\section{The $K$-User Gaussian Interference Channel with Confidential Messages}
\label{sec:ConfidentialMessages}

To illustrate the main idea, we start with the intuitive argument
for the three-user Gaussian interference channel. Let $F=2m+1$ for
some $m\in\NN$. This is the $(2m+1)$ symbol extension of the
three-user channel considered in~\cite{Cadembe:Interference}. We now
employ interference alignment precoding using the matrices
$\bar{\Vm}_k$ of~\cite{Cadembe:Interference}, so that the
transmitted signals are of the form $\bar{\Xm}_k (t) = \bar{\Vm}_k
\tilde{\Xm}_k (t) $, where $\tilde{\Xm}_k (t)$ represents the vector
of $m_k$ streams transmitted from user $k$ (see Fig.~$3$). According
to the interference alignment principle, the beamforming matrices
$\bar{\Vm}_k$ are constructed to satisfy the following two
properties:

$1$) The non-intended signals seen by each receiver are aligned
within some low dimensionality subspace. More precisely, the column
space of the matrices $\Hm_{i,k}\bar{\Vm}_k$ for $k\in\Kc-i$ lie in
a subspace of dimension $F-m_i$ at receiver $i$.

$2$) The intended streams span the orthogonal subspace, i.e., the
columns of $\Hm_{i,i}\bar{\Vm}_i$ are independent and are orthogonal
to that of $\Hm_{i,k}\bar{\Vm}_k$ for each user $k\in\Kc-i$.

This way, the $F$ dimensional received signal space at each receiver
is used to create $m_i$ interference free dimensions, spanned by the
desired streams. Now, let us consider receiver $1$ as the
eavesdropper for the messages of users $2$ and $3$. This particular
eavesdropper now sees the $m$ streams $\tilde{\Xm}_2 (t)$ and $m$
streams $\tilde{\Xm}_3 (t)$ mixed together in a multiple access
channel with {\bf only} $m$ dimensions. This key observation allows
for the secrecy precoding $\tilde{\Xm}_2(t)$ and $\tilde{\Xm}_3(t)$
to {\em completely secure} $m/2$ streams in each transmitted vector.
It is easy to see that a similar argument follows for securing each
vector against the second potential eavesdropper. In the limit of a
large $F=2m+1$, the $m/2$ secure streams results in $1/4$ secure
DoF. This intuitive discussion is formalize for the general case of
a $K$-user Gaussian interference channel in the following.

\begin{theorem}\label{thm:KUserConf}
For the $K$-user Gaussian interference channel with confidential
messages, a secure DoF of $\eta=\frac{K-2}{2K-2}$ per frequency-time
slot is almost surely achievable for each user.
\end{theorem}

\begin{proof}
We shall show that almost all codebooks in an appropriately
constructed ensemble satisfy the achievability conditions for
symmetric secure DoF of $\eta=\frac{K-2}{2K-2}$ with probability
that approaches to $1$, for all channel coefficients, as
$n,m,\rho\to\infty$.

Fix an $m\in \NN$. Let $m_1=(m+1)^M$, $m_k=m^M$ $\forall k\neq 1$,
$M=(K-1)(K-2)-1$, and $F=(m+1)^M+m^M$ frequency slots. We now
generate, for each user $k$,
$2^{nm_k\left(\frac{F}{m_k}(R_k+R_k^x)\right)}$ codewords each of
length $nm_k$ with entries that are independent and identically
distributed (i.i.d.) $\sim
\Cc\Nc\left(0,\frac{\rho-\epsilon}{c_k}\right)$. We choose $c_k$ to
satisfy the power constraint for each user:
$c_k=\frac{tr(\bar{\Vm}_k\bar{\Vm}_k^H)}{F}$. These codewords are
then randomly partitioned into $M_k=2^{nFR_k}$ message bins, each
consisting of $M_k^x=2^{nFR_k^x}$ codewords. Hence, an entry of the
$k^{th}$ user codebook will be represented by
$\hat{\Xm}_k(w_k,w_k^x)$ where the bin index $w_k\in \Wc_k$ is the
secrecy message and the index $w_k^x\in \{1,\cdots,M_k^x\}$ is the
randomization message. It is easy to see that the secure
transmission rate per orthogonal time and frequency slot is $R_k$.

To send a message $w_k$, the $k^{th}$ transmitter looks into the bin
$w_k\in \Wc_k$ and randomly selects a codeword in this bin, denoted
by the index $w_k^x$, according to uniform distribution. It thus
obtains $\hat{\Xm}_k(w_k,w_k^x)$ of length $nm_k$. We further
partition the elements of this vector as
$\hat{\Xm}_k(w_k,w_k^x)=[\tilde{\Xm}_k(1),\cdots,\tilde{\Xm}_k(n)]$,
where each element is an $m_k \times 1$ vector. Then, for each
symbol time $t \in\{1,\cdots,n\}$, the transmitter employs the
interference alignment scheme, and maps $\tilde{\Xm}_k(t)$ to
$\bar{\Xm}_k(t)$ via $\bar{\Xm}_k(t)=\bar{\Vm}_k\tilde{\Xm}_k(t)$,
where we choose the interference alignment matrices $\bar{\Vm}_k$ as
in~\cite{Cadembe:Interference}.

We choose the secrecy and randomization rates as
follows~\footnote{Since the channel coefficients are fixed and known
everywhere, we omit the conditioning on them here.}.

\begin{eqnarray}\label{eq:Rchoice}
R_k&=&\frac{1}{F} \min_{i\in\Kc}\left\{
I(\tilde{\Xm}_i;\bar{\Ym}_i) \right\}
-
\frac{1}{(K-1)F} \max_{i\in\Kc}
\left\{ I(\tilde{\Xm}_{\Kc-i};\bar{\Ym}_i) \right\}\quad \text{and} \nonumber\\
R_k^x&=&\frac{1}{F} \min_{i\in\Kc, \Sc\subseteq\Kc-i}
\left\{ \frac{1}{|\Sc|}
I(\tilde{\Xm}_{\Sc};\bar{\Ym}_i|\tilde{\Xm}_{\Kc-\Sc-i} )
\right\}
\end{eqnarray}

The above rates are inside the decodability region for each user,
i.e., $R_k+R_k^x \leq \frac{1}{F} I(\tilde{\Xm}_k;\bar{\Ym}_k),$
$\forall k\in\Kc$, implying that each user can reliably decode its
own streams as $n\to\infty$. Hence, using the union bound argument,
we can show that for a given $\epsilon$ there exists $n_0(\epsilon)$
such that for any $n>n_0(\epsilon)$
$\max\{P_{e,1},\cdots,P_{e,K}\}\leq \epsilon$ for almost all
codebooks in the ensemble. Our second step is to show that
$\Delta_{\Sc,i}$ can be made arbitrarily close to $1$ for any
$i\in\Kc$ and $\Sc\subset\Kc-i$ for almost all codebooks in the
ensemble. Towards this end, it is sufficient to focus on the
equivocation at an arbitrary receiver $i\in\Kc$. Furthermore, it is
sufficient to establish perfect secrecy for the full message set
since Lemma~\ref{thm:secrecy} shows that perfect secrecy of the full
message set implies secrecy for all subsets (Here, the full message
set at the receiver $i$ refers to $W_{\Kc-i}$.). Denoting the
observation of the eavesdropper as $\Ym_i$, we write
\begin{eqnarray}\label{eq:entropies}
H(W_{\Kc-i}|\Ym_i) &=& H(W_{\Kc-i},\Ym_i) - H(\Ym_i) \nonumber\\
&=& H(W_{\Kc-i},W_{\Kc-i}^x,\Ym_i)
- H(W_{\Kc-i}^x|W_{\Kc-i},\Ym_i)- H(\Ym_i) \nonumber\\
&=& H(W_{\Kc-i})+ H(W_{\Kc-i}^x|W_{\Kc-i})
+ H(\Ym_i|W_{\Kc-i},W_{\Kc-i}^x) \\
&&{-}\: H(W_{\Kc-i}^x|W_{\Kc-i},\Ym_k)- H(\Ym_i) \nonumber\\
&=& H(W_{\Kc-i})+H(W_{\Kc-i}^x)-I(W_{\Kc-i},W_{\Kc-i}^x;\Ym_i)
-H(W_{\Kc-i}^x|W_{\Kc-i},\Ym_i),\nonumber
\end{eqnarray}
where the last equality follows from the fact that
$H(W_{\Kc-i}^x|W_{\Kc-i})= H(W_{\Kc-i}^x)$ as the randomization
(i.e., codeword) indices are independent of the message (i.e., bin)
indices. We now bound each term of~\eqref{eq:entropies}. First
\begin{equation}
I(W_{\Kc-i},W_{\Kc-i}^x;\Ym_i)\leq
I(\tilde{\Xm}_{\Kc-i}(1),\cdots,\tilde{\Xm}_{\Kc-i}(n);\Ym_i)
\end{equation}
due to the Markov chain relationship
\begin{equation}
\{W_{\Kc-i},W_{\Kc-i}^x \}\to
\{\tilde{\Xm}_{\Kc-i}(1),\cdots,\tilde{\Xm}_{\Kc-i}(n)\} \to \Ym_i.
\end{equation}
Combining this with the fact that
\begin{equation}
I(\tilde{\Xm}_{\Kc-i}(1),\cdots,\tilde{\Xm}_{\Kc-i}(n);\Ym_i) \leq n
\max\limits_{p(\tilde{\Xm}_{\Kc-i})}
I(\tilde{\Xm}_{\Kc-i};\bar{\Ym}_i), \nonumber
\end{equation}
we obtain
\begin{equation}\label{eq:bound1}
I(W_{\Kc-i},W_{\Kc-i}^x;\Ym_i)\leq n
\max\limits_{p(\tilde{\Xm}_{\Kc-i})}
I(\tilde{\Xm}_{\Kc-i};\bar{\Ym}_i).
\end{equation}

Second
\begin{eqnarray}\label{eq:bound2}
H(W_{\Kc-i}^x)=\log\left(\prod\limits_{k\neq i}M_k^x\right)
=nF\sum\limits_{k\in\Kc-i} R_k^x.
\end{eqnarray}

To upper bound the last term, we use the following argument. Assume
that $w_{\Kc-i}\in\Wc_{\Kc-i}$ is transmitted. Given these bin
indices, the remaining randomness in $W_{\Kc-i}^x$ at the
eavesdropper can be resolved for almost all codebooks as the above
choice of $R_k^x$ satisfies the multiple access channel
achievability conditions $\sum\limits_{k\in\Sc}R_k^x \leq
\frac{1}{F}
I(\tilde{\Xm}_{\Sc};\bar{\Ym}_i|\tilde{\Xm}_{\Kc-\Sc-i})$, $\forall
\Sc\subset\Kc-i$ ~\cite[Chapter 14]{ThomasAndCover}. Then, by Fano's
inequality, we have $H(W_{\Kc-i}^x|W_{\Kc-i}=w_{\Kc-i},\Ym_i) \leq
n\delta(n,w_{\Kc-i})$, where $\delta(n,w_{\Kc-i})\to 0$ as
$n\to\infty$. Then, defining $\delta(n)\triangleq
\max\limits_{w_{\Kc-i}\in \Wc_{\Kc-i}} \delta(n,w_{\Kc-i})$, we have
\begin{eqnarray}\label{eq:bound3}
H(W_{\Kc-i}^x|W_{\Kc-i},\Ym_i)&=&
\sum\limits_{w_{\Kc-i}\in\Wc_{\Kc-i}} H(W_{\Kc-i}^x|W_{\Kc-i}=w_{\Kc-i},\Ym_i)
\:p(W_{\Kc-i}=w_{\Kc-i})\nonumber\\
&\leq&  n \delta(n),
\end{eqnarray}
where $\delta(n)\to 0$ as $n\to\infty$. Using equations
(\ref{eq:bound1}), (\ref{eq:bound2}), and (\ref{eq:bound3}) in
(\ref{eq:entropies}) and dividing both sides of by $H(W_{\Kc-i})$,
we obtain
\begin{eqnarray}\label{eq:equivocation}
\Delta_{\Kc-i,i} \geq 1 - \hat{\delta},
\end{eqnarray}
\begin{eqnarray}\label{eq:deltahat1}
\hat{\delta} &\triangleq& \frac{\delta(n)
+  \max\limits_{p(\tilde{\Xm}_{\Kc-i})}
I(\tilde{\Xm}_{\Kc-i};\bar{\Ym}_i)
- F\sum\limits_{k\in\Kc-i} R_k^x  }{F\sum\limits_{k\in\Kc-i} R_k}, \nonumber\\
&=&\frac{\delta(n)
+  \max\limits_{p(\tilde{\Xm}_{\Kc-i})}
I(\tilde{\Xm}_{\Kc-i};\bar{\Ym}_i)
- (K-1) \min\limits_{i\in\Kc, \Sc\subseteq\Kc-i}
\left\{ \frac{1}{|\Sc|}
I(\tilde{\Xm}_{\Sc} ;\bar{\Ym}_i |\tilde{\Xm}_{\Kc-\Sc-i}  )
\right\} }
{(K-1) \min\limits_{i\in\Kc}\left\{
I(\tilde{\Xm}_i ;\bar{\Ym}_i ) \right\}
-
\max\limits_{i\in\Kc}
\left\{ I(\tilde{\Xm}_{\Kc-i} ;\bar{\Ym}_i ) \right\}},
\end{eqnarray}
where we used the fact that
$H(W_{\Kc-i})=nF\sum\limits_{k\in\Kc-i} R_k$ and the
rate assignment given by \eqref{eq:Rchoice}.

It is already observed that $\delta(n)\to 0$ as $n\to\infty$ for
almost all codebooks in the ensemble. The orthogonality of the
intended message and interference at each respective receiver along
with the full rank property of the gain matrices (see
Lemma~\ref{thm:LemmaFullRank}) imply the followings.

\begin{equation}\label{eq:ProofKUserConfEq1}
\lim_{\rho\to\infty}\frac{\max\limits_{p(\tilde{\Xm}_{\Kc-i})}
I(\tilde{\Xm}_{\Kc-i};\bar{\Ym}_i)}{ \log(\rho)}
=F-m_i,
\: \forall i\in\Kc,
\end{equation}
\begin{equation}\label{eq:ProofKUserConfEq2}
\lim_{\rho\to\infty}
\frac{I(\tilde{\Xm}_{\Sc};\bar{\Ym}_i|\tilde{\Xm}_{\Kc-\Sc-i})}{ \log(\rho)}
=r,
\end{equation}
where $r=m^M$ or $r=(m+1)^M$ depending on $i$,
\begin{equation}\label{eq:ProofKUserConfEq3}
\lim_{\rho\to\infty}\frac{I(\tilde{\Xm}_i;\bar{\Ym}_i)}{ \log(\rho)}
=m_i,
\: \forall i\in\Kc,
\end{equation}
and
\begin{equation}\label{eq:ProofKUserConfEq4}
\lim_{\rho\to\infty}\frac{I(\tilde{\Xm}_{\Kc-i};\bar{\Ym}_i)}{ \log(\rho)}
=F-m_i,
\: \forall i\in\Kc.
\end{equation}

Using the observations \eqref{eq:ProofKUserConfEq1}, \eqref{eq:ProofKUserConfEq2},
\eqref{eq:ProofKUserConfEq3}, and \eqref{eq:ProofKUserConfEq4} in
\eqref{eq:deltahat1} we see that
\begin{equation}
\lim\limits_{n,m,\rho\to\infty} \hat{\delta} = 0
\end{equation}
for almost all codebooks in the ensemble. Hence, for any given
$\epsilon>0$, we can make $\Delta_{\Kc-i,i} \geq 1-\epsilon$ by
letting $n,m,\rho$ grow. Finally, due to \eqref{eq:Rchoice},
\eqref{eq:ProofKUserConfEq3}, and \eqref{eq:ProofKUserConfEq4}, we
obtain
\begin{equation}
\eta=\lim\limits_{m,\rho\to\infty}
\frac{R_k}{ \log(\rho)}=\frac{K-2}{2K-2}.
\end{equation}
which proves our result.
\end{proof}


\section{The $K$-User Gaussian Interference Channel with an External Eavesdropper}
\label{sec:ExternalEavesdropper}

First, it is easy to see that our previous results extend naturally
when the eavesdropper CSI is available {\em a-priori} at the
different transmitters and receivers. Intuitively, one can imagine
the existence of a virtual transmitter associated with the external
eavesdropper transforming our $K$-user network into another one with
$K+1$-users. This way, one can achieve a secure DoF of
$\eta=\frac{(K+1)-2}{2(K+1)-2}=\frac{K-1}{2K}$ per frequency-time
slot for each user using the scheme of the previous section. For
example, for a two-user network with an external eavesdropper, it is
possible to achieve $\frac{1}{4}$ secure DoFs if the eavesdropper
CSI is available at the transmitters. More formally, we have the
following result.
\begin{corollary}
For the $K$-user Gaussian interference channel with an external
eavesdropper, a secure DoF of $\eta=\frac{K-1}{2K}$ per
frequency-time slot is almost surely achievable for each user
(assuming the availability of the eavesdropper CSI).
\end{corollary}

More interestingly, it is still possible to achieve positive secure
DoF per user in the {\bf absence} of the eavesdropper CSI by
exploiting the channel {\bf ergodicity}. In the ergodic model, the
channel gains are assumed to be fixed during a block of $n_1$ symbol
times and then randomly change to another value in the next block
for a total of $B$ blocks, where $n_1\rightarrow\infty$ and
$B\rightarrow\infty$.

Again, for illustration purposes, we use the $K=3$ case. Here, the
users of the network have $\frac{3m+1}{2m+1}$ total DoF while the
multiple access channel (MAC) seen by the eavesdropper can have only
one DoF from its observations. Hence, via an appropriate choice of
secrecy codebooks, the $\frac{m}{2m+1}$ additional DoF can be
\emph{evenly} distributed among the network users {\em on the
average}, allowing for a $\frac{1}{6}$ secure DoF per user without
any requirement on the eavesdropper CSI. In the general case, we
have the following result.

\begin{theorem}\label{thm:KUserExt}
For the $K$-user Gaussian interference channel with an external
eavesdropper, a secure DoF of
$\eta=\frac{1}{2}-\frac{1}{K}$ per frequency-time slot is
achievable for each user in the ergodic
setting (in the absence of the eavesdropper CSI).
\end{theorem}

\begin{proof}
Let $m\in\NN$ and $F=(m+1)^M+m^M$, where $M=(K-1)(K-2)-1$. We set
$m_1=(m+1)^M$ and $m_k=m^M$ for $k\neq 1$. We generate all the
permutations of length $K$ and denote this set by $\Pi$, where
$|\Pi|=K!$. Then, for each fading block $b\in\{1,\cdots,B\}$, we
randomly pick, according to uniform distribution, a permutation from
$\Pi$ and denote it by $\pi_b$. In order to ensure statistical
symmetry, the interference alignment matrices in each fading block
will be obtained according to a different user ordering induced by
$\pi_b$. More specifically, let $k(b)=\pi_b(k)$ and
$\Hm_{i(b),k(b)}^{(b)}=\Hm_{i,k}(b)$. Using the newly ordered
channel matrices $\Hm_{i(b),k(b)}^{(b)}$, the interference alignment
matrix for the user $k(b)$, i.e., $\bar{\Vm}_{k(b)}$, is generated.

For each secrecy codebook in the ensemble, we generate $2^{n F
(R_k+R_k^x)}$ sequences each of length $n_1\sum\limits_{b=1}^B
m_{k(b)}$, where entries are chosen i.i.d. $\sim \Cc\Nc
\left(0,\frac{P-\epsilon}{c}\right)$ for some $\epsilon>0$ and $c$
that satisfies the long term average power constraint (the existence
of $\epsilon$ and $c$ follows from the argument of
Theorem~\ref{thm:KUserConf}). We independently assign each codeword
to one of $M_k=2^{n F R_k}$ bins each having $M_k^x=2^{n F R_k^x}$
codewords. Given $w_k$, transmitter $k$ chooses the corresponding
bin and independently (according to uniform distribution) chooses a
codeword in that bin denoted by $\hat{\Xm}_k(w_k,w_k^x)$, where
$w_k^x$ is the randomization index. This codeword is then divided
into $B$ blocks, each with a length $n_1m_{k(b)}$ symbols. Each
block is then arranged in the following $m_{k(b)}\times n_1$ matrix
$[\tilde{\Xm}_k(1+(b-1)n_1),\cdots ,\tilde{\Xm}_k(n_1+(b-1)n_1)]$,
where $\tilde{\Xm}_k(j+(b-1)n_1)$, for $1\leq j\leq n_1$, is an
$m_{k(b)}\times 1$ vector. At time slot $t=j + (b-1)n_1$, the
$k^{th}$ transmitter maps $\tilde{\Xm}_k(t)$ to $\bar{\Xm}_k(t)$ via
$\bar{\Xm}_k(t)=\bar{\Vm}_{k(b)} \tilde{\Xm}_k(t)$. Finally, we
would like to emphasize the fact that in the sequel expectations
will be taken with respect to the random distribution of the channel
matrices and the uniform distribution underlying the permutation
operators used in different fading blocks.

Our first key observation is that the equivalent channel matrices
$\Hm_{i,k}(b)\bar{\Vm}_{k(b)}$ connecting $\tilde{\Xm}_k(t)$ and
$\bar{\Ym}_i(t)$ are identically distributed $\forall i,k\in\Kc$ and
$b\in\{1,\cdots,B\}$. This property will allow us to drop the
subscript $i$ and write
$\mathbb{E}[I(\tilde{\Xm}_i;\bar{\Ym}_i|\Hm)] =
\mathbb{E}[I(\tilde{\Xm};\bar{\Ym}|\Hm)] $, $\forall i\in\Kc$ in the
following. To satisfy the achievability conditions and the secrecy
requirements of the network, we choose $R_k$ and $R_k^x$ as follows
\begin{eqnarray}\label{eq:ErgodicRkchoice}
R_k &=& \frac{1}{KF} \left(K
\mathbb{E}[I(\tilde{\Xm};\bar{\Ym}|\Hm)]-
\max\limits_{p(\bar{\Xm}_{\Kc})}\mathbb{E}
[I(\bar{\Xm}_\Kc;\bar{\Ym}_e|\Hm,\Hm_e)]\right)\quad \text{and} \nonumber\\
R_k^x &=&\frac{1}{KF} \mathbb{E}[I(\tilde{\Xm}_{\Kc};\bar{\Ym}_e|\Hm,\Hm_e)],
\end{eqnarray}
where the maximization in the first equation is among all possible
input distributions. With this choice of rates, we have the
following
\begin{eqnarray}
R_k+R_k^x
&=&\frac{1}{F}
\mathbb{E}[I(\tilde{\Xm};\bar{\Ym}|\Hm)]-
\frac{1}{KF}\max\limits_{p(\bar{\Xm}_{\Kc})}
\mathbb{E}[I(\bar{\Xm}_\Kc;\bar{\Ym}_e|\Hm,\Hm_e)]
+ \frac{1}{KF} \mathbb{E}[I(\tilde{\Xm}_{\Kc};\bar{\Ym}_e|\Hm,\Hm_e)]\nonumber\\
&\leq& \frac{1}{F}
\mathbb{E}[I(\tilde{\Xm};\bar{\Ym}|\Hm)],
\end{eqnarray}
where the inequality is due to the maximization among \emph{all}
possible input distributions in the second term of the equation.
Hence, we have $R_k+R_k^x\leq
\frac{1}{F}\mathbb{E}[I(\tilde{\Xm};\bar{\Ym}|\Hm)]$, from which we
conclude that each user in the interference network can decode its
own secrecy and randomization indices as $n_1\to\infty$ and as
$B\to\infty$ (using almost all codebooks in the ensemble). The next
step is to study the equivocation at the eavesdropper, i.e.,
\begin{eqnarray}\label{eq:entropies2}
\frac{1}{n}H(W_{\Kc}|\Ym_e,\Hm,\Hm_e) &=& \frac{1}{n}\big(H(W_{\Kc},\Ym_e,\Hm,\Hm_e) - H(\Ym_e,\Hm,\Hm_e) \big) \nonumber\\
&=& \frac{1}{n}\big(H(W_{\Kc},W_{\Kc}^x,\Ym_e,\Hm,\Hm_e)
- H(W_{\Kc}^x|W_{\Kc},\Ym_e,\Hm,\Hm_e)\nonumber\\
&&{-} \:H(\Ym_e,\Hm,\Hm_e) \big) \nonumber\\
&=& \frac{1}{n}\big(H(W_{\Kc})+ H(W_{\Kc}^x|W_{\Kc})
+ H(\Ym_e,\Hm,\Hm_e|W_{\Kc},W_{\Kc}^x) \nonumber\\
&&{-} \: H(W_{\Kc}^x|W_{\Kc},\Ym_e,\Hm,\Hm_e)- H(\Ym_e,\Hm,\Hm_e) \big) \nonumber\\
&=& \frac{1}{n}\big( H(W_{\Kc})+H(W_{\Kc}^x)-I(W_{\Kc},W_{\Kc}^x;\Ym_e,\Hm,\Hm_e)\nonumber\\
&&{-} \: H(W_{\Kc}^x|W_{\Kc},\Ym_e,\Hm,\Hm_e) \big) ,
\end{eqnarray}
where the last equality follows from the fact that
$H(W_{\Kc}^x|W_{\Kc})= H(W_{\Kc}^x)$ as the codeword indices are
independent of the bin indices. Here,
\begin{eqnarray}
H(W_\Kc^x) = \log \left(\prod\limits_{k=1}^K M_k^x\right) =
\sum\limits_{k=1}^K n F R_k^x = n
\mathbb{E}[I(\tilde{\Xm}_{\Kc};\bar{\Ym}_e|\Hm,\Hm_e)]
\end{eqnarray}
and
\begin{eqnarray}
\lim\limits_{n\to\infty}\frac{1}{n}I(W_\Kc,W_\Kc^x;\Ym_e,\Hm,\Hm_e) &\leq &
\lim\limits_{n\to\infty}\frac{1}{n}I(\tilde{\Xm}_{\Kc}(1),\cdots,\tilde{\Xm}_{\Kc}(n);\Ym_e,\Hm,\Hm_e)\nonumber\\
&=&\lim\limits_{n\to\infty}\frac{1}{n}\big(I(\tilde{\Xm}_{\Kc}(1),\cdots,\tilde{\Xm}_{\Kc}(n);\Hm,\Hm_e)\nonumber\\
&&\:{+}I(\tilde{\Xm}_{\Kc}(1),\cdots,\tilde{\Xm}_{\Kc}(n);\Ym_e|\Hm,\Hm_e)\big)\nonumber\\
&=&\lim\limits_{n\to\infty}\frac{1}{n}I(\tilde{\Xm}_{\Kc}(1),\cdots,\tilde{\Xm}_{\Kc}(n);\Ym_e|\Hm,\Hm_e)\nonumber\\
&=& \mathbb{E}[I(\tilde{\Xm}_{\Kc};\bar{\Ym}_e|\Hm,\Hm_e)],
\end{eqnarray}
where the first inequality is due to the Markov chain relationship
$$\{W_\Kc,W_\Kc^x\} \to
\{\tilde{\Xm}_{\Kc}(1),\cdots,\tilde{\Xm}_{\Kc}(n)\}\to
\{\Ym_e,\Hm,\Hm_e\},$$
and the last one is due to ergodicity. For the
last term of (\ref{eq:entropies2}), we observe that the channel seen
by the eavesdropper reduces to a fading MAC channel for the
randomization messages due to the code construction. For this fading
MAC, each user is able to set its randomization message rate as a
fraction $\frac{1}{K}$ of the total DoF seen by the eavesdropper as
chosen in~(\ref{eq:ErgodicRkchoice}), and assure the decodability of
the randomization messages at the eavesdropper given the secrecy
message indices (the technical details are reported in
Lemma~\ref{thm:SimpleLemma1}, Lemma~\ref{thm:SimpleLemma2}, and
Lemma~\ref{thm:RandRates}). Then, by Fano's inequality, we have
$$\lim\limits_{n_1,B\to\infty}
\frac{H(W_\Kc^x | W_\Kc,\Ym_e,\Hm,\Hm_e)}{n_1B} = 0,$$ for almost
all codebooks in the ensemble. Therefore, by dividing both sides of
(\ref{eq:entropies2}) by $\frac{1}{n}H(W_\Kc)$, we can ensure
\begin{equation}
\Delta_{\Kc}=\frac{H(W_\Kc|\Ym_e,\Hm,\Hm_e)}
{H(W_\Kc)} \geq 1- \epsilon,
\end{equation}
for any $\epsilon>0$ as $n_1,B\to\infty$, which is sufficient for
our purposes (please refer to Lemma~\ref{thm:secrecy}). Finally,
considering (\ref{eq:ErgodicRkchoice}), we have
$$\lim\limits_{\rho\to\infty}\frac{\mathbb{E}[I(\tilde{\Xm} ;\bar{\Ym}| \Hm )]}
{\log(\rho)}=
\left(\frac{1}{K}{m_1}+\frac{K-1}{K}{m_2}\right), $$
and hence
\begin{eqnarray}
\lim\limits_{\rho\to\infty}\frac{R_k}{\log(\rho)}
&=&\frac{1}{KF}\left(K \left(\frac{1}{K}{m_1}+\frac{K-1}{K}{m_2}\right)-
F\right) \nonumber\\
&=& \frac{(K-2)m^M}{KF}
\end{eqnarray}
implying that $\eta=\frac{m^M}{F}- \frac{2m^M}{KF}$ DoF is
achievable for each user for any $m$. Consequently, we conclude that
$\lim\limits_{m\to\infty}\eta=\frac{1}{2}- \frac{1}{K}$ symmetric
DoF is achievable with perfect secrecy in the ergodic setting.

\end{proof}

It is important to observe that the achievability of a positive DoF
for the no eavesdropper CSI scenario hinges largely on the
ergodicity assumption, whereas when the eavesdropper CSI is assumed
to be available our results hold almost surely for all channel
realizations. This is the price entailed by the lack of eavesdropper
CSI. Finally, the positive impact of interference on the secrecy
capacity region is best illustrated by comparing our results to the
point-to-point scenario. In~\cite{Gopala:On}, a point-to-point
channel with an external eavesdropper was shown to have zero DoF. On
the other hand, our results show that as more source-destination
pairs are added to the network, each pair is able to achieve
non-zero DoF for $K\geq2$. This seemingly surprising result is due
to interference alignment technique which {\bf not only} allows for
a clean separation between the intended message and interference at
each receiver, but also packs the interfering signals into a low
dimensionality subspace, and hence, impairs the ability of each
eavesdropper to distinguish any of the secure messages efficiently.


\section{Conclusions}
\label{sec:Conclusion}

In this work, we have considered the $K$-user Gaussian interference
channel with secrecy constraints. By using the interference
alignment scheme along with secrecy pre-coding at each transmitter,
we have shown that each user in the network can achieve a non-zero
secure DoF. Our results differentiate between the confidential
messages scenario and the case where an external eavesdropper, with
unknown CSI, is present in the network. The most interesting aspect
of our results is, perhaps, the discovery of the role of
interference in increasing the secrecy capacity of multi-user
wireless networks.


\appendices

\section{Lemma~\ref{thm:secrecy}}
\label{sec:Proofsecrecy}

\begin{lemma}\label{thm:secrecy}
Consider receiver $i\in\Kc$. For a given $\epsilon>0$ and $d \in[0,1]$,
$\exists \epsilon^*(i,\epsilon,d)>0$ such that,
if $\Delta_{\Kc-i,i}\geq 1-\epsilon^*(i,\epsilon,d)$ then
$\Delta_{\Sc,i}\geq d-\epsilon$, $\forall \Sc\subseteq\Kc-i$.
\end{lemma}

\begin{proof}
For a given  $i\in\Kc$, $\epsilon>0$, and level of secrecy $d \in[0,1]$,
let $\epsilon^*(i,\epsilon,d)=\min\limits_{\Sc\subseteq\Kc-i}(1+\epsilon-d)
\frac{H(\Wc_{\Sc})}{H(\Wc_{\Kc-i})}$. Then,
denoting the received observation of the eavesdropper as $\Ym_i$
and assuming $\Delta_{\Kc-i,i}\geq 1-\epsilon^*(i,\epsilon,d)$,
for any $\Sc\subseteq \Kc-i$ we have
\begin{IEEEeqnarray}{l}\label{eq:lemma1ineq}
H(\Wc_{\Sc}|\Ym_i) + H(\Wc_{\Kc-i}|\Wc_{\Sc},\Ym_i)
= H(\Wc_{\Kc-i}|\Ym_i) \nonumber\\
\geq H(\Wc_{\Kc-i})- \epsilon^*(i,\epsilon,d) H(\Wc_{\Kc-i}) \\
\geq H(\Wc_{\Sc})+H(\Wc_{\Kc-i}|\Wc_{\Sc})
- (1+\epsilon-d)H(\Wc_{\Sc}),\nonumber
\end{IEEEeqnarray}
where the first inequality follows from the assumption
of $\Delta_{\Kc-i,i}\geq 1-\epsilon^*(i,\epsilon,d)$ and the second inequality follows
from the choice of $\epsilon^*(i,\epsilon,d)$ above. Continuing from above,
\begin{IEEEeqnarray}{l}
\Delta_{\Sc,i}  = \frac{H(\Wc_{\Sc}|\Ym_i)}{H(\Wc_{\Sc})}
\geq (d-\epsilon)
+ \frac{H(\Wc_{\Kc-i}|\Wc_{\Sc})
-H(\Wc_{\Kc-i}|\Wc_{\Sc},\Ym_i)}{H(\Wc_{\Sc})}
\geq (d-\epsilon),\nonumber
\end{IEEEeqnarray}
as conditioning does not increase entropy.
\end{proof}

\section{Lemma~\ref{thm:LemmaFullRank}}
\label{sec:ProofLemmaFullRank}

\begin{lemma}\label{thm:LemmaFullRank}
The gain matrix, resulting from the interference alignment scheme,
between transmitter $k$ and the receiver $i$, i.e.,
$\Hm_{i,k}\bar{\Vm}_k$, has rank $m_k$ with probability one. As the
dimension of $\Hm_{i,k}\bar{\Vm}_k$ is $F \times m_k$, these
matrices have full rank with probability one.
\end{lemma}

\begin{proof}
We have $\textrm{rank}(\Hm_{k,k}\bar{\Vm}_k)=m_k$ by the
construction given in~\cite{Cadembe:Interference}. Now, the second
observation follows by the design of interference alignment
vectors, which have linearly independent columns (If they had
linearly dependent columns, then $\Hm_{k,k}\bar{\Vm}_k$ would not
have $m_k$ linearly independent columns, contrary to the
construction of the interference alignment matrices.). Here,
$\textrm{rank}(\Hm_{i,k}\bar{\Vm}_k)\leq\min\{m_k,F\}=m_k$ for $i
\neq k$. We need only to show that the matrix
$\Hm_{i,k}\bar{\Vm}_k$ has $m_k$ linearly independent columns.
Considering any $i \neq k$, representing diagonal elements of
$\Hm_{i,k}$ as $\{h_{i,k}(1),h_{i,k}(2),\cdots,h_{i,k}(F)\}$ and
denoting the rows of the interference alignment matrix by $\vv_f$,
i.e., $\bar{\Vm}_k=[\vv_1^T; \vv_2^T; \cdots ; \vv_F^T]^T$, we
have $\Hm_{i,k}\bar{\Vm}_k=[h_{i,k}(1)\vv_1^T; h_{i,k}(2)\vv_2^T;
\cdots; h_{i,k}(F)\vv_F^T]^T$. At this point, as the channel gains
are chosen according to a continuous distribution, the
$h_{ik}(f)$'s are non-zero with probability one for
$f\in\{1,2,\cdots,F\}$. Hence, these row operations will not
change the rank of a matrix, i.e.,
$\textrm{rank}(\Hm_{i,k}\bar{\Vm}_k)=\textrm{rank}(\bar{\Vm}_k)=m_k$.
Therefore, the gain matrices seen by the receivers have full rank
with probability one.
\end{proof}

\section{Lemma~\ref{thm:SimpleLemma1}}
\label{sec:ProofSimpleLemma1}

\begin{lemma}\label{thm:SimpleLemma1}
For any $\Mc,\Lc \subset \Kc$ satisfying $\Mc\cap\Lc=\emptyset$,
\begin{eqnarray}
I(\tilde{\Xm}_{\Mc};\bar{\Ym}_e|\Hm,\Hm_e)
\leq I(\tilde{\Xm}_{\Mc};\bar{\Ym}_e|\tilde{\Xm}_{\Lc},\Hm,\Hm_e).\nonumber
\end{eqnarray}
\end{lemma}

\begin{proof}
\begin{eqnarray}
I(\tilde{\Xm}_{\Mc} ;\bar{\Ym}_e |\Hm,\Hm_e)&=&H(\tilde{\Xm}_{\Mc} |\Hm,\Hm_e)
-H(\tilde{\Xm}_{\Mc} |\bar{\Ym}_e ,\Hm,\Hm_e)\nonumber\\
&\leq& H(\tilde{\Xm}_{\Mc} |\tilde{\Xm}_{\Lc} ,\Hm,\Hm_e)-
H(\tilde{\Xm}_{\Mc} |\bar{\Ym}_e ,\tilde{\Xm}_{\Lc} ,\Hm,\Hm_e)\\
&=& I(\tilde{\Xm}_{\Mc} ;\bar{\Ym}_e |\tilde{\Xm}_{\Lc} ,\Hm,\Hm_e),\nonumber
\end{eqnarray}
where the inequality is due to the fact that conditioning
does not increase entropy, and the last equality follows by
$H(\tilde{\Xm}_{\Mc} |\tilde{\Xm}_{\Lc},\Hm,\Hm_e)=H(\tilde{\Xm}_{\Mc}|\Hm,\Hm_e)$
as $\Mc\cap\Lc=\emptyset$ and
messages of the users are independent.
\end{proof}

\section{Lemma~\ref{thm:SimpleLemma2}}
\label{sec:ProofSimpleLemma2}

\begin{lemma}\label{thm:SimpleLemma2}
\begin{eqnarray}
\frac{1}{|\Sc^c|}\mathbb{E}[I(\tilde{\Xm}_{\Sc^c} ;\bar{\Ym}_e |\Hm,\Hm_e)]
\leq \frac{1}{|\Sc|}
\mathbb{E}[I(\tilde{\Xm}_{\Sc} ;\bar{\Ym}_e |\tilde{\Xm}_{\Sc^c} ,\Hm,\Hm_e)]
\end{eqnarray}
\end{lemma}

\begin{proof}
Let us denote $|\Sc|=S$; and define
$\Sc=\{s_1,\cdots,s_S\}$ and
$\Sc^c=\{s_{S+1},\cdots,s_K\}$. Then we have
\begin{IEEEeqnarray}{l}
\frac{1}{|\Sc^c|}\mathbb{E}[I(\tilde{\Xm}_{\Sc^c} ;\bar{\Ym}_e |\Hm,\Hm_e)]\nonumber\\
=\frac{1}{K-S}\bigg(\mathbb{E}[I(\tilde{\Xm}_{s_{S+1}} ;\bar{\Ym}_e |\Hm,\Hm_e)]+
\mathbb{E}[I(\tilde{\Xm}_{s_{S+2}} ;\bar{\Ym}_e |\tilde{\Xm}_{s_{S+1}} ,\Hm,\Hm_e)]
+\cdots\nonumber\\
\: \: \: \: \: \: {+}\:
\mathbb{E}[I(\tilde{\Xm}_{s_K} ;\bar{\Ym}_e |\tilde{\Xm}_{s_{S+1}} ,\cdots,
\tilde{\Xm}_{s_{K-1}} ,\Hm,\Hm_e)] \bigg)\nonumber\\
\leq\frac{1}{K-S}\bigg(\mathbb{E}
[I(\tilde{\Xm}_{s_1} ;\bar{\Ym}_e |\tilde{\Xm}_{\Sc^c} ,\Hm,\Hm_e)]+
\mathbb{E}[I(\tilde{\Xm}_{s_1} ;\bar{\Ym}_e |\tilde{\Xm}_{\Sc^c},\Hm,\Hm_e)]
+\cdots\nonumber\\
\: \: \: \: \: \: {+}\:
\mathbb{E}[I(\tilde{\Xm}_{s_1} ;\bar{\Ym}_e |\tilde{\Xm}_{\Sc^c},\Hm,\Hm_e )] \bigg)\nonumber\\
=\frac{1}{K-S}\left((K-S)\mathbb{E}
[I(\tilde{\Xm}_{s_1} ;\bar{\Ym}_e |\tilde{\Xm}_{\Sc^c},\Hm,\Hm_e )]\right)\nonumber\\
=\frac{1}{S}\left(S \mathbb{E}
[I(\tilde{\Xm}_{s_1} ;\bar{\Ym}_e |\tilde{\Xm}_{\Sc^c},\Hm,\Hm_e )]\right)\nonumber\\
=\frac{1}{S}\bigg(\mathbb{E}
[I(\tilde{\Xm}_{s_1} ;\bar{\Ym}_e |\tilde{\Xm}_{\Sc^c},\Hm,\Hm_e )]+
\mathbb{E}[I(\tilde{\Xm}_{s_2} ;\bar{\Ym}_e |\tilde{\Xm}_{\Sc^c},\Hm,\Hm_e )]
+\cdots\nonumber\\
\: \: \: \: \: \: {+}\:
\mathbb{E}[I(\tilde{\Xm}_{s_{S}} ;\bar{\Ym}_e |\tilde{\Xm}_{\Sc^c},\Hm,\Hm_e )] \bigg)\nonumber\\
\leq\frac{1}{S}\bigg(\mathbb{E}
[I(\tilde{\Xm}_{s_1} ;\bar{\Ym}_e |\tilde{\Xm}_{\Sc^c},\Hm,\Hm_e )]+
\mathbb{E}[I(\tilde{\Xm}_{s_2} ;
\bar{\Ym}_e |\tilde{\Xm}_{\Sc^c} ,\tilde{\Xm}_{s_1},\Hm,\Hm_e )]+\cdots\nonumber\\
\: \: \: \: \: \: {+}\:
\mathbb{E}[I(\tilde{\Xm}_{s_{S}} ;\bar{\Ym}_e |
\tilde{\Xm}_{\Sc^c} ,\tilde{\Xm}_{s_1} ,\cdots,
\tilde{\Xm}_{s_{S-1}},\Hm,\Hm_e )] \bigg)\nonumber\\
= \frac{1}{|\Sc|}
\mathbb{E}[I(\tilde{\Xm}_{\Sc} ;\bar{\Ym}_e |\tilde{\Xm}_{\Sc^c},\Hm,\Hm_e )],
\end{IEEEeqnarray}
where we repeatedly use Lemma~\ref{thm:SimpleLemma1} for
inequalities and use the fact that
$\mathbb{E}[I(\tilde{\Xm}_{k} ;\bar{\Ym}_e |\tilde{\Xm}_{\Lc},\Hm,\Hm_e )]
=\mathbb{E}[I(\tilde{\Xm}_{i} ;\bar{\Ym}_e |\tilde{\Xm}_{\Lc},\Hm,\Hm_e )]$ for
any $k\neq i$ and for any $\Lc\subset\Kc-\{k,i\}$. We note that
the last property stated above is due to the symmetry between
network users provided by the random choice of user ordering at
each fading block.

\end{proof}

\section{Lemma~\ref{thm:RandRates}}
\label{sec:ProofRandRates}

\begin{lemma}\label{thm:RandRates}
Each user can set the randomization rates to be
$\frac{1}{K}$th of the total DoF per orthogonal
time-frequency slot seen by the eavesdropper, i.e.,
with a rate choice of
\begin{eqnarray}
R_k^x
=\frac{1}{KF} \mathbb{E}[I(\tilde{\Xm}_{\Kc} ;\bar{\Ym}_e|\Hm,\Hm_e )],
\end{eqnarray}
each randomization message (codeword index), given the secrecy
message (bin index) of each user, is decodable at the eavesdropper.
\end{lemma}

\begin{proof}
Let $\Sc\subset\Kc$.
From Lemma~\ref{thm:SimpleLemma2}, we have
$$\frac{1}{|\Sc^c|}\mathbb{E}[I(\tilde{\Xm}_{\Sc^c} ;\bar{\Ym}_e|\Hm,\Hm_e )]
\leq \frac{1}{|\Sc|}
\mathbb{E}[I(\tilde{\Xm}_{\Sc} ;\bar{\Ym}_e |\tilde{\Xm}_{\Sc^c},\Hm,\Hm_e )].$$
We continue as below.
\begin{eqnarray}
\frac{1}{|\Sc^c|}\mathbb{E}[I(\tilde{\Xm}_{\Sc^c} ;\bar{\Ym}_e|\Hm,\Hm_e )]
&\leq& \frac{1}{|\Sc|} \mathbb{E}
[I(\tilde{\Xm}_{\Sc} ;\bar{\Ym}_e |\tilde{\Xm}_{\Sc^c},\Hm,\Hm_e )]\\
\Leftrightarrow
|\Sc|\mathbb{E}[I(\tilde{\Xm}_{\Sc^c} ;\bar{\Ym}_e|\Hm,\Hm_e )]
&\leq& (K-|\Sc|) \mathbb{E}
[I(\tilde{\Xm}_{\Sc} ;\bar{\Ym}_e |\tilde{\Xm}_{\Sc^c},\Hm,\Hm_e )]\\
\Leftrightarrow \frac{|\Sc|}{K}
\mathbb{E}[I(\tilde{\Xm}_{\Sc^c} ;\bar{\Ym}_e|\Hm,\Hm_e )]
&\leq& \frac{K-|\Sc|}{K} \mathbb{E}
[I(\tilde{\Xm}_{\Sc} ;\bar{\Ym}_e |\tilde{\Xm}_{\Sc^c},\Hm,\Hm_e )]\\
\Leftrightarrow \frac{|\Sc|}{K}
\mathbb{E}[I(\tilde{\Xm}_{\Kc} ;\bar{\Ym}_e|\Hm,\Hm_e )] &\leq&
\mathbb{E}[I(\tilde{\Xm}_{\Sc} ;\bar{\Ym}_e |\tilde{\Xm}_{\Sc^c},\Hm,\Hm_e )],
\end{eqnarray}
from which we readily conclude that
$R_k^x=\frac{1}{KF}\mathbb{E}[I(\tilde{\Xm}_{\Kc} ;\bar{\Ym}_e|\Hm,\Hm_e )]$
satisfies
\begin{eqnarray}
\sum\limits_{k\in\Sc} R_k^x=
\frac{|\Sc|}{KF} \mathbb{E}[I(\tilde{\Xm}_{\Kc} ;\bar{\Ym}_e|\Hm,\Hm_e )]
\leq \frac{1}{F}\mathbb{E}
[I(\tilde{\Xm}_{\Sc} ;\bar{\Ym}_e |\tilde{\Xm}_{\Sc^c},\Hm,\Hm_e )],
\forall \Sc\subset\Kc,
\end{eqnarray}
and hence randomization messages are decodable at
the eavesdropper with this rate assignment.
\end{proof}


\newpage


\newpage

\begin{figure}[htb] 
    \centering
    \includegraphics[width=0.8\columnwidth]{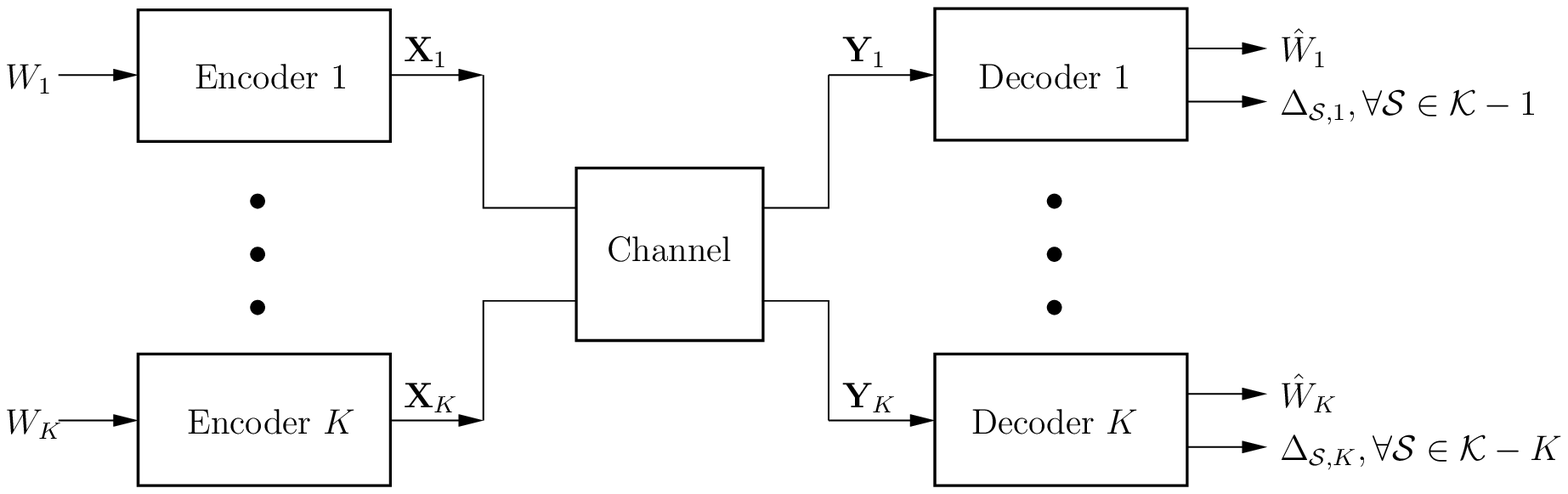}
    \caption{$K$-user interference channel with confidential messages.}
\end{figure}

\begin{figure}[htb] 
    \centering
    \includegraphics[width=0.8\columnwidth]{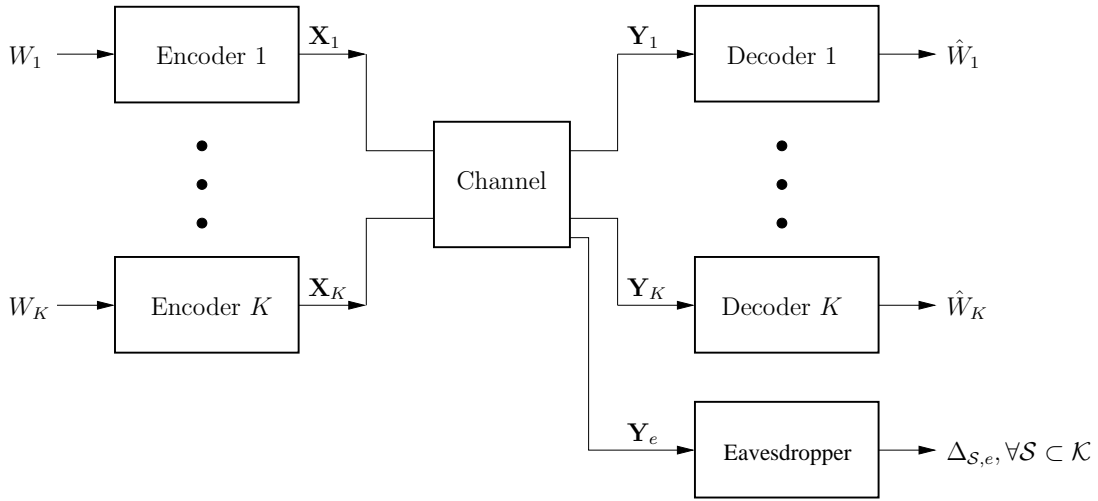}
    \caption{$K$-user interference channel with an external eavesdropper.}
\end{figure}

\begin{figure}[htb] 
    \centering
    \includegraphics[width=0.6\columnwidth]{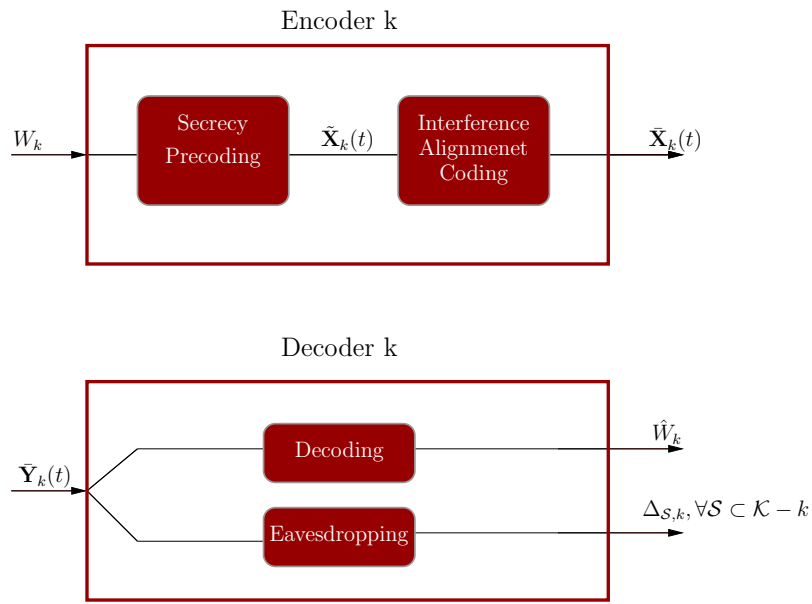}
    \caption{Proposed encoder and decoder architecture for user $k$
    in the $K$-user interference channel with confidential messages.}
\end{figure}

\begin{figure}[htb] 
    \centering
    \includegraphics[width=0.5\columnwidth]{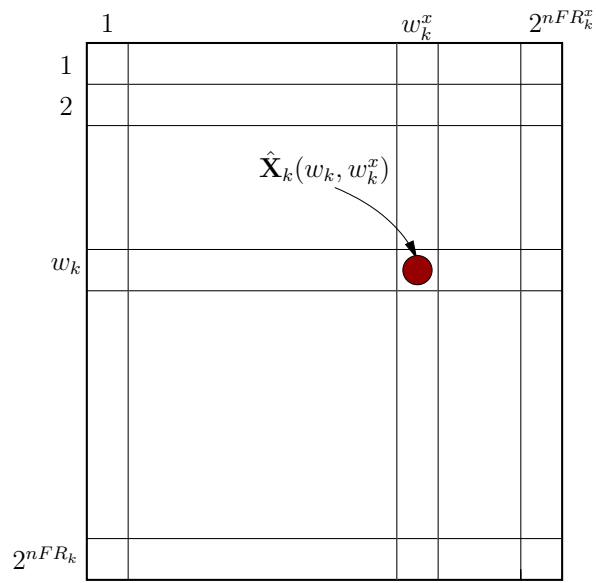}
    \caption{Proposed codebook structure for user $k$. The secret
    message of user $k$ results in the bin index $w_k$ and the randomization
    index for the corresponding user is $w_k^x$. Considering these two
    indices, each entry of the codebook is denoted as $\hat{\Xm}_k(w_k,w_k^x)$.}
\end{figure}

\end{document}